\documentclass{ifacconf}
\usepackage{xcolor}
\usepackage{xfrac}
\usepackage{amsfonts}
\usepackage{amsmath}
\usepackage[cal=boondox]{mathalfa}
\usepackage{mathtools}
\usepackage{tikz}
\usepackage{graphicx}      
\usepackage{natbib}        

\newtheorem{lemma}{Lemma}

\begin{document}
\begin{frontmatter}

   \title{Maneuvering-based Dynamic Thrust Allocation for Fully-Actuated Vessels}

   \thanks[footnoteinfo]{Work funded by the Research Council of Norway (RCN) through NTNU AMOS (RCN project 223254), SFI AutoShip (RCN project 309230), and the Polish National Centre for Research and Development through the ENDURE project (NOR/POLNOR/ENDURE/0019/2019-00).}

   \author[First]{Emir Cem Gezer}
   \author[First]{Roger Skjetne}

   \address[First]{Department of Marine Technology, Norwegian University of Science and Technology, NO-7491 Trondheim, Norway (e-mail: [emir.cem.gezer,roger.skjetne]@ntnu.no).}

   \begin{abstract}
      This paper introduces a new approach to solving the thrust allocation problem using the maneuvering problem in the maritime domain for fully actuated vessels.
      The method uses a control Lyapunov function to create a nonlinear reference filter for the thruster forces.
      The filter ensures dynamic tracking of the optimal thrust allocation solution with rate limitation in the output thruster references.
      It further uses control barrier functions to ensure that the thruster force saturation limits are respected.
      The approach aims for simplicity and effectiveness, as well as smooth and dynamic thruster reference signals, in the implementation of thrust allocation for marine vessels.
   \end{abstract}


\end{frontmatter}

\section{Introduction} \label{sec:introduction}

Most maneuvering tasks mainly aim to guide an object along a desired path; additional objectives, such as speed and acceleration during the maneuver, may not be essential.
There are several approaches for solving the maneuvering objective, and the most prominent ones are tracking control and path following control.
The \textit{tracking} controller forces the object to follow a defined path by ensuring the system output $y(t) \in \mathbb{R}^m$ closely tracks a desired output $y_d(t)$ that traces out the intended trajectory, whereas the \textit{path-following} controller guides the object onto and along a predefined time-independent path without necessarily considering the time scheduling or speed assignment along the path \citep{fossen2011handbook}.
With this in mind, the formulation of the \textit{maneuvering problem} \citep{skjetne2005maneuvering} aims to bridge the gap between the tracking and path-following control by allowing secondary objectives to be considered through the use of an additional \textit{path parameter}.
It separates the maneuvering into two tasks: a \textit{geometric task} and a \textit{dynamic task}.
Here, the objective of the geometric task is to force the object towards a desired parameterized path, while the dynamic task focuses on the particulars of the maneuver using the \textit{path parameter}.
The maneuvering problem builds upon the theories provided by \citet{hauser1995maneuver} and \citet{encarnaccao2001combined}, and further example applications of such control designs are presented in \citep{aguiar2008performance} and \citep{peng2016containment}.
The generic maneuvering problem \citep{skjetne2011line} is an abstraction of the classical maneuvering problem, where the aim now is more generally to maneuver a dynamical system (not necessarily a vessel) according to a geometric task and a dynamic task.
In this formulation, the desired path is rephrased into a desired manifold  in the state space, parametrized by a vector manifold parameter.
As such, the classical maneuvering problem becomes a special case; see also \citet{skjetne2005maneuvering}, Ch. 3.5.

The \textit{thrust allocation}, as per the application in this paper, determines the reference inputs for the thrusters to ensure a desired resultant control load is induced on the vessel.
A wide variety of marine vehicles, such as dynamic positioning (DP) vessels, autonomous underwater vehicles (AUVs), and remotely operated vehicles (ROVs), use thrust allocation, and requirements for each application vary.
For example, ROVs are often only equipped with fixed thrusters, whereas DP vessels use a combination of fixed and azimuth thrusters where there are numerous considerations to make when computing the azimuth angles to account for certain cases such as biasing mode and forbidden sectors.
Owing to the nature of the problem, it is often formulated as a \textit{constrained optimization problem}.
Thus, approaches from numerical optimization such as linear programming (LP), quadratic programming (QP), and sequential quadratic programming (SQP), etc., are considered as state-of-the-art methods \citep{johansen2013control}.

This work proposes a method to dynamically solve the thrust allocation problem as a \textit{generic maneuvering problem}.
The outcome becomes a reference filter that computes the desired thruster forces by using a variable $\theta$ that parameterizes the nullspace of the thruster configuration matrix, which enables us to consider additional control objectives.
The dynamics of the reference filter are based on a Control Lyapunov Function (CLF) to solve the geometric task and a Control Barrier Function (CBF) to handle thruster saturation limits.
The dynamic task is designed to minimize a cost function $J(t, \theta)$ to ensure an optimal behavior for the thrusters.

The rest of the paper is structured as follows.
In Section \ref{sec:problem_formulation} we lay out the problem and describe and introduce the formulations relevant to our method.
Then, we present our method to solve both geometric and dynamic tasks in Section \ref{sec:method}.
We present two different case studies where the maneuvering-based control allocation approach was used in Section \ref{sec:case_studies}.
We show our preliminary results in Section \ref{sec:results}, and then present our conclusions in Section \ref{sec:conclusions}.

\section{Problem Formulation}\label{sec:problem_formulation}

\subsection{Background}

The thruster configuration of a marine vessel with $m$ thrusters is defined by a set of lever arm coordinate vectors $l_i \in \mathbb{R}^r, \ i \in \{1, \dots, m\}$.
These vectors represent the displacement from the vessel's coordinate origin (CO) to the location of each thruster.
Here, $r \in \mathbb{N}$ denotes the number of dimensions in the workspace; for instance, $r=2$ for planar motion and $r=3$ for three-dimensional motion.
For a single thruster, let $F_i \in \mathbb{R}_{\geq 0}$ be the force magnitude, $f_i \in \mathbb{R}^r$ be the force vector in each axis that the thruster $i$ exerts, and the force and moment vector (we will refer to this as the \textit{load vector}) it produces in the body frame be denoted by $\tau_i \in \mathbb{R}^n$, where $n$ is the number of supported degrees of freedom.
Then, $\tau = \sum_{i=1}^{m}\tau_i, \ \tau \in \mathbb{R}^n$, represents the total thrust load exerted by $m$ thrusters in $r$-dimensional space.
Each thruster's contribution, $\tau_i \in \mathbb{R}^n$, can be formulated by either the rectangular thrust configuration matrix (sometimes referred to as extended thrust configuration) $B_i \in \mathbb{R}^{n \times r}$ or the polar matrix $B_i(\alpha_i) \in \mathbb{R}^{n \times 1}$, where $\alpha_i$ is the azimuth angle.
Notably, some thrusters may have a fixed angle $\alpha_i$, while azimuth thrusters vary their direction.
For the latter, separating the force components and using the rectangular configuration matrix is convenient as this leads to a constant configuration matrix.

Consider the following setup of a horizontal 3-DOF (surge, sway, yaw) marine vessel motion, so that $n=3$.
Let there be $m_1$ varying-azimuth thrusters and $m_2$ fixed-direction thrusters so that $m=m_1+m_2$.
We will now only represent the varying-azimuth thrusters in rectangular coordinate formulation, whereas fixed-azimuth thrusters are kept with their polar coordinate formulation.
In this example, the force vector of the $i^{\text{th}}$ thruster becomes $f_i = [X_i, Y_i]^\top \in \mathbb{R}^2$ and the force magnitude is $F_i \in \mathbb{R}_{\geq 0}, \ F_i = |f_i|$.
Say $B_1 \in \mathbb{R}^{n\times 2m_1}$ is used for rectangular thrust allocation matrix and $B_2 \in \mathbb{R}^{n \times 2 m_2}$ for the polar thrust allocation matrix, and let $\mathbb{S}^1 \coloneqq \{x \in \mathbb{R}^2 : x^\top x = 1 \}$ be the unit circle, $a_i \coloneqq [\cos \alpha_i , \sin \alpha_i]^\top, \ a_i \in \mathbb{S}^1$ is the azimuth direction vector, and $S\coloneqq \begin{bmatrix}0 & -1 \\ 1 & 0\end{bmatrix}$.
We can then formulate the total thrust load according to
\begin{align}
    & \tau = \sum_{i=1}^{m_1} B_{1,i} f_i + \sum_{i=m_1 + 1}^{m} B_{2,i} F_i \label{eq:01_mixed_thrust_allocation_general} \\
    & B_{1,i} \coloneqq \begin{bmatrix} I \\ l_i^\top S^\top \end{bmatrix}, \
   B_{2,i} \coloneqq \begin{bmatrix} a_i \\ l_i^\top S^\top a_i \end{bmatrix}. \label{eq:01_mixed_thrust_allocation_expanded}
\end{align}
The relationship between force magnitude and the azimuth angle is $\alpha_i \coloneqq \angle f_i = \angle a_i$ for the varying azimuth thruster.
We can further simplify the notation by defining the thrust configuration matrix $B \in \mathbb{R}^{n \times p}$, where $p \coloneqq 2m_1 + m_2$.
Then, we can write the total thrust load effector model as
\begin{align}
   \tau = B \xi, \quad B \coloneqq \begin{bmatrix}B_1 & B_2\end{bmatrix} \in \mathbb{R}^{n \times p}
\end{align}
where $\xi \coloneqq \text{col}(f_1, \dots, f_{m_1}, F_{m_1 + 1}, \dots, F_m) \in \mathbb{R}^p$ is the vector of thruster forces, and $\xi_i$ is the corresponding force for the $i^\text{th}$ thruster.
Note that the $\xi_i$ component represents either $f_i$ or $F_i$, and its dimension (2 or 1) thus differs to reflect the thruster type.
From here, the problem is to solve the thrust allocation problem where $\tau$ dynamically tracks a commanded load input $\tau_{cmd}$.

\section{Method}\label{sec:method}

We will apply the \textit{Maneuvering Problem} as the method for solving our problem.
The reader may refer to the generic maneuvering approach reported by \cite{skjetne2011line} for a detailed explanation.

\subsection{Geometric Task}

Let $\tau_{cmd}$ be a commanded thrust load, assuming $\tau_{cmd}$ and $\dot{\tau}_{cmd}$ are provided externally, and we require $\text{rank}(B) = n$ and $p > n$.
Our reference model will be given by
\begin{align}
   \dot{\xi}_i = \phi_i, \ i \in \{1,\dots,m\}, \ \tau = B \xi \label{eq:03_control_input_phi},
\end{align}
where $ \xi \coloneqq \text{col}(\xi_1, \xi_2, \dots, \xi_m) \in \mathbb{R}^p$, and $B \in \mathbb{R}^{n \times p}$.
The overall objective in the design of our reference filter is to design control laws for $\phi_i$ to ensure that
\begin{align}
   \lim_{t \to \infty} | \tau(t) - \tau_{cmd}(t) | = 0.
\end{align}
To this end, we design a desired $\xi_d$ so that
\begin{align}
   B \xi_d = \tau_{cmd}
   \label{eq:02_design_ensures},
\end{align}
where we let $\xi_d = \xi_p + \xi_0$ with $\xi_p$ a particular solution to \eqref{eq:02_design_ensures} and $\xi_0 \in \mathcal{N}(B)$, where $\mathcal{N}(B)$ represents the nullspace of $B$.
Given a $\xi_p$ satisfying $B\xi_p (t) = \tau_{cmd}(t)$, we can use $\xi_0$ to explore other possible solutions within $\mathcal{N}(B)$.
Let $q \coloneqq p - n$ be the dimension of the nullspace, and let the columns of $Q \in \mathbb{R}^{p \times q}$ form an orthonormal basis for it.
The nullspace can then be parametrized by $\theta \in \mathbb{R}^q$ according to
\begin{align}
   \begin{aligned}
      \mathcal{N}(B) \coloneqq \{\xi_0 \in \mathbb{R}^p : \xi_0 = Q\theta, \ \theta \in \mathbb{R}^q \}.
   \end{aligned}
\end{align}
This gives the following $q$-dimensional time-varying manifold describing the solution space of \eqref{eq:02_design_ensures},
\begin{align}
   \begin{aligned}
      \xi_d(t, \theta) \coloneqq \xi_p(t) + Q \theta
   \end{aligned}.
   \label{eq:02_nullspace}
\end{align}
The geometric task is then to design $\phi_i$ to ensure that
\begin{align}
   \lim_{t \to \infty} | \xi(t) - \xi_d(t, \theta(t)) | = 0.
   \label{02:eq_geometric_task_xi}
\end{align}

\subsection{Dynamic Task}

The objective of the dynamic assignment will be to ensure that $\theta$ moves to the minimizer of a cost function $J(t, \theta)$, where $\theta \mapsto J(t,\theta)$ is convex $\forall t \geq 0$.
For continuous-time systems, convergence to the minimizer of $J$ can be obtained by a gradient method.
Hence, we define the dynamic assignment
\begin{align}
   \upsilon(t, \theta) \coloneqq - \gamma \nabla_{\theta}J(t, \theta)^\top,
   \label{eq:03_dynamic_assignment}
\end{align}
where $\gamma > 0$ is the descent gain.
Hence, for the dynamic task, we will direct the parameterization variable $\theta$ so that
\begin{equation}
   \lim_{t \to \infty} |\dot{\theta}(t) - \upsilon(t, \theta(t))| = 0.
   \label{eq:03_dynamic_task}
\end{equation}

\subsection{Nominal Maneuvering Control Design}

We now solve the control allocation as a maneuvering problem with a dynamic control design where $\xi \in \mathbb{R}^p$ is considered a virtual force state in the reference model with dynamics \eqref{eq:03_control_input_phi}.
This is next used to generate the desired individual signals for the local thruster controllers.

We use a CLF to design the nominal control law $\phi_i = \kappa_i$,
\begin{align}
   V(t, \theta, \xi) \coloneqq \sum_{i=1}^{m}\frac{c_i}{2}\tilde{\xi_i}^\top \tilde{\xi_i},
   \quad
   \tilde{\xi}_i \coloneqq \xi_i - \xi_{d,i}(t,\theta),
   \label{eq:03_clf_00}
\end{align}
where $c_i > 0$ is a relative weight between the thrusters.
The total time derivative of $V(t,\theta, \xi)$ becomes
\begin{align}
   \dot{V} = \sum \limits_{i=1}^m c_i \tilde{\xi_i} ^\top  \left[ \phi_i - \frac{\partial \xi_{d,i}(t,\theta)}{\partial t} - \frac{\partial \xi_{d,i}(t, \theta)}{\partial \theta} \dot{\theta} \right].
   \label{eq:02_clf}
\end{align}
Since $\frac{\partial \xi_{d,i}(t, \theta)}{\partial t} = \dot{\xi}_{p,i}(t)$ and $\frac{\partial \xi_{d,i}(t, \theta)}{\partial \theta} = Q_i$, where $Q_i$ is the rows of $Q$ corresponding to the $i^\text{th}$ thruster, we can choose
\begin{align}
   \phi_i = \kappa_i(t, \theta, \xi_i) \coloneqq -\bar{\Omega}_i \frac{\tilde{\xi_i}}{|\tilde{\xi_i}| + \zeta_i} + \dot{\xi}_{p,i}(t) + Q_i\upsilon(t, \theta), 
   \label{eq:03_nominal_control_law}
\end{align}
where the saturation parameter $\bar{\Omega}_i > 0$ is set to the rate limit of the $i^\text{th}$ thruster, and $\zeta_i$ shapes the slope of the nonlinear feedback at zero.
Finally, note that
\begin{align}
   \frac{\partial V(t, \theta, \xi)}{\partial \theta} = \sum_{i=1}^{m} c_i \tilde{\xi}_i^\top \frac{\partial \tilde{\xi}_i(t, \theta)}{\partial \theta} = -\sum_{i=1}^{m} c_i \tilde{\xi}_i^\top Q_i.
   \label{eq:03_nominal_maneuvering_control_design}
\end{align}
Hence, choosing
\begin{align}
   \dot{\theta} = \upsilon(t, \theta) - \mu \Bigg[\frac{\partial V(t, \theta, \xi)}{\partial \theta}\Bigg]^\top , \quad \mu \geq 0,
   \label{eq:02_manuevering_control_law}
\end{align}
completes the nominal design, where $\mu$ is recognized as the common maneuvering gradient gain in \citep{skjetne2005maneuvering}.

\subsection{Safeguarding control design to respect constraints}

The nominal control law is designed without considering the actuator saturation limits.
Yet, a control allocation algorithm must obey the limitations of the actuators.
To this end, we use a CBF approach to ensure the control signals stay within the saturation limits.
For each thruster with dynamics \eqref{eq:03_control_input_phi}, we have designed \eqref{eq:03_nominal_control_law} as the nominal controls.
Next, we embed this into a safety layer where each force vector $\xi_i$ must be limited to respect the saturation constraint $|\xi_i|\leq F_{i,\text{max}}$.
Correspondingly, we define the CBF $C_i(\xi_i)$ and the safe set $S_i$ according to
\begin{gather}
   C_i(\xi_i) \coloneqq \xi_i^\top \xi_i - F_{i,\text{max}}^2, \label{eq:02_cbf_0} \\
   S_i \coloneqq \{\xi_i \in \mathbb{R}^k: C_i(\xi_i) \leq 0\}, \label{eq:02_cbf_safe_set}
\end{gather}
where $S_i$ must be rendered forward invariant.
The total time derivative of $C_i$ becomes
\begin{align}
   \dot{C}_i = 2\xi_i^\top \dot{\xi}_i = 2 \xi_i^\top \phi_i,
\end{align}
and we then identify the following set of safe controls, i.e.,
\begin{align}
   \begin{aligned}
      U_i(\xi_i) \coloneqq & \{\phi_i \in \mathbb{R}^k : \dot{C}_i(\xi_i) \leq - \frac{1}{\rho_i} C_i(\xi_i)\}                         \\
      =                    & \{\phi_i \in \mathbb{R}^k : \xi_i^\top \xi_i - F^2_{i, \text{max}} + 2 \rho_i \xi_i^\top \phi_i  \leq 0\}
   \end{aligned}
\end{align}
where $\rho_i > 0$ is a time constant for tuning, and $k \in \{1, 2\}$ depending on the dimension of $\xi_i$.
The following lemma will aid in assigning the safe control inputs.

\begin{lemma}
   \label{lemma:cbf_lemma}
   Given a safe state set $K \subset \mathbb{R}^\mathcal{n}$, suppose a corresponding safe control set $\mathcal{U}(x) \coloneqq \{u \in \mathbb{R}^\mathcal{m} : a(x) + b(x)^\top u \leq 0\}$, where $a : \mathbb{R}^\mathcal{n} \rightarrow \mathbb{R}$ and $b : \mathbb{R}^\mathcal{n} \rightarrow \mathbb{R}^\mathcal{m}$ are continuously differentiable functions.
   Let $\kappa_0(x)$ be nominal control, and assume $b(x) \neq 0$, $\forall x \in \{x \in \mathbb{R}^\mathcal{n} : \kappa_0(x) \notin \mathcal{U}(x)\}$.
   The optimal control problem
   \begin{align}
      \kappa^*(x) = & \arg \min_{u \in \mathcal{U}(x)}(u -\kappa_0(x))^\top \Gamma (u-\kappa_0(x))
   \end{align}
   with $\Gamma = \Gamma^\top > 0$, has the solution
   \begin{align}
      \kappa^* \coloneqq
      \begin{cases}
         \kappa_0,                                                               & \kappa_0 \in \mathcal{U}    \\
         \kappa_0 - \frac{a + b^\top \kappa_0}{b^\top \Gamma^{-1}b}\Gamma^{-1}b, & \kappa_0 \notin \mathcal{U}
      \end{cases}
   \end{align}
   where function arguments $(x)$ were left out for brevity.
\end{lemma}
The proof follows from the work by e.g. \citet{xu2015robustness}. See also \citep{marley2021maneuvering}.

To get the safe control inputs with the CBF given in \eqref{eq:02_cbf_0}, we apply Lemma \ref{lemma:cbf_lemma}, using
\begin{align}
   a \coloneqq \xi_i^\top \xi_i - F^2_{i, \text{max}}, \quad b \coloneqq 2 \rho_i \xi_i, \quad  \Gamma \coloneqq I,
   \label{eq:03_cbf_1}
\end{align}
and obtain the following optimal control which respects the saturation limits,
\begin{align}
   \begin{aligned}
      \phi_i = \kappa_{i,\text{safe}} (t, \theta,\xi_i) \coloneqq
      \begin{cases}
         \kappa_i,                                                   & \kappa_i \in U_i(\xi_i)     \\
         \kappa_i - \frac{a_i + b_i^\top \kappa_i}{b_i^\top b_i}b_i, & \kappa_i \notin U_i(\xi_i).
      \end{cases}
   \end{aligned}
   \label{eq:03_cbf_2}
\end{align}

Alternatively, the following CBF can be used,
\begin{gather}
   \begin{gathered}
      C_i(\xi_i) \coloneqq |\xi_i| - F_{i,\text{max}}, \quad \dot{C}_i = \frac{\xi_i^\top}{|\xi_i|}\phi_i
      \label{eq:03_cbf_abs_1}
   \end{gathered} \\
   \begin{gathered}U_i = \Bigg\{ \phi_i \in \mathbb{R}^k :  |\xi_i| - F_{i,\text{max}} + \rho_i\frac{\xi_i^\top}{|\xi_i|} \phi_i \leq 0 \Bigg\},
      \label{eq:03_cbf_abs_2}
   \end{gathered}
\end{gather}
giving a linear force measure to the saturation limit, which may be preferable over the quadratic CBF \eqref{eq:02_cbf_0}.

\subsection{The resulting reference filter}

The reference filter takes the desired load vector $\tau_{cmd}$ and its derivative $\dot{\tau}_{cmd}$ as inputs and makes the filter virtual force state $\xi$ track the desired parameterized state $\xi_d(t, \theta)$.
The state $\xi$ is then used to calculate the respective outputs of the reference filter, that is, the allocated magnitudes $F_{i}$ and the allocated directions $\alpha_{i}$ for the thrusters.
For each application, a cost function $J(t, \theta)$ relevant to the dynamic task needs to be designed and embedded into the nominal controls $\kappa_i$.
The reference filter then applies the safeguarding control presented in \eqref{eq:03_nominal_control_law}, \eqref{eq:03_cbf_1}, and \eqref{eq:03_cbf_2} to avoid exceeding the force saturation limits.
\begin{align}
   \tau_{cmd}, \dot{\tau}_{cmd} \rightarrow
   \boxed{
      \text{Reference Filter}(\xi, \theta)
   } \rightarrow
   {F}_{i}, \alpha_{i}
\end{align}
where $F_i := |\xi_i|$ and $\alpha_i := \angle \xi_i$ for the azimuth thrusters.
Note that in scenarios where the derivative of $\tau_{cmd}$ is not readily available, it can be approximated using a lowpass filter.
In other cases, when the rate of change in $\tau_{cmd}$ is slow compared to the bandwidth of the local thruster control systems, it may be sufficient to set $\dot{\tau}_{cmd}$ to zero (which is normally the case for DP control systems).
\section{Case Studies}\label{sec:case_studies}

In the following case studies, we first solve the unconstrained control allocation problem where the goal is to find a particular solution $\xi_p$ that satisfies the desired load vector $\tau_{cmd}$.
We then assign the dynamic task and handle the saturations.
As a particular solution, the unconstrained minimum 2-norm solution can be applied by using the weighted Moore-Penrose pseudoinverse,
\begin{gather}
   \begin{gathered}
      B^\dagger_W \coloneqq W^{-1}B^\top [BW^{-1} B^\top]^{-1}, \\
      \xi_p(t) \coloneqq B^{\dagger}_W \tau_{cmd}(t),
   \end{gathered}
   \label{eq:02_weighted_control_allocation_solution}
\end{gather}
where $W \coloneqq \text{diag}(w_1 I,\dots,w_{m_1} I, w_{m_1+1},\dots,w_m) > 0$ is a weight matrix that allows prioritization between thrusters.

\subsection{Minimum squared norm thrust allocation}

Consider the minimum magnitude thrust allocation problem where the aim is to keep the thruster forces as small as possible.
We define a consistent convex cost function $J(t, \theta)$ for the dynamic task according to
\begin{align}
   K(z) \coloneqq \frac{1}{2} z^\top W z, \quad  J(t, \theta) \coloneqq K(\xi_{d}(t, \theta)),
   \label{eq:04_potential_func_k}
\end{align}
to obtain the optimal nullspace parameter $\theta^*$.
Using the steepest descent method, we apply the gradient
\begin{align}
   \nabla_\theta J(t, \theta) =
   \xi_d(t, \theta)^\top W \frac{\partial \xi_d(t, \theta)}{\partial \theta} = \xi_d(t, \theta)^\top WQ
\end{align}
in \eqref{eq:03_dynamic_assignment}.
The $\text{argmin}_\theta J(t, \theta)$ can now explicitly be found by calculating the gradient and setting it to zero, i.e.,
\begin{align}
   \begin{aligned}
      0 = \nabla_\theta J(t, \theta^*)^\top & = Q^\top W \xi_d(t, \theta^*)                               \\
                                            & = Q^\top W (B^\dagger_W \tau_{cmd}(t) + Q \theta^*)         \\
                                            & = Q^\top W B_W^\dagger \tau_{cmd}(t) + Q^\top W Q \theta^*.
   \end{aligned}
\end{align}
Since $Q$ forms the basis for $\mathcal{N}(B)$, we get $Q^\top W B_W^\dagger = 0$.
Furthermore, $Q$ being full rank implies $\theta^* = 0$.
Hence, the global minimum of $J(t, \theta)$ for each $t \geq 0$ corresponds to $\xi_d(t, \theta^*) = B_W^\dagger \tau_{cmd}(t)$ as expected.

To complete the dynamic task, we recall \eqref{eq:03_dynamic_assignment} and obtain
\begin{align}
   \upsilon(t, \theta) = -\gamma Q^\top W Q \theta
   \label{eq:03_dynamic_assignment_00}
\end{align}
with $\gamma > 0$ being the descent gain.
Choosing $c_i=w_i$ for the CLF in \eqref{eq:03_clf_00}, and assigning this to \eqref{eq:03_nominal_maneuvering_control_design}, we conclude the \textit{minimum squared norm reference filter} thrust allocation using \eqref{eq:03_dynamic_assignment_00} as
the dynamic task in \eqref{eq:02_manuevering_control_law}.

\begin{figure}
   \centering
   \includegraphics[width=0.8\linewidth]{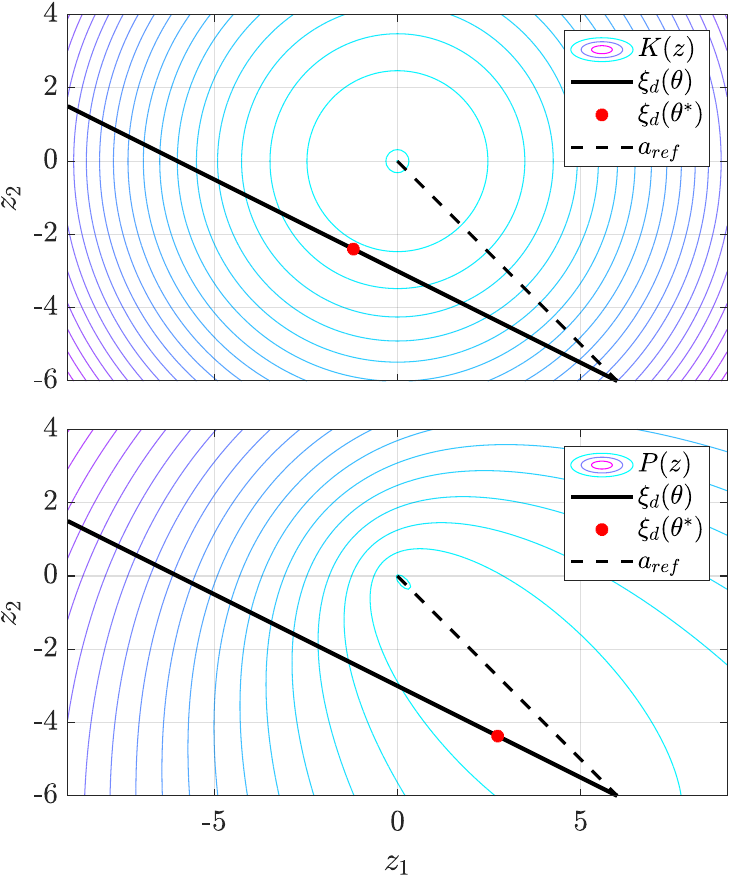}
   \caption{Contour plot of $K(z)$ (top) and $P(z)$ (bottom), with $B=[1, 2]$, $\tau_{cmd} = -6$, $z \in \mathbb{R}^2$, $a_{ref} = [\sqrt{2}, - \sqrt{2}]^\top$, and the black solid line indicating $\mathcal{N}(B)$.}
   \label{fig:stacked_cost_function_j1}
\end{figure}

As a simple example, we have plotted the contour lines of $z \mapsto K(z)$ with $W=I$, $B=[1,2]$, $\tau_{cmd}=-6$ so that $n=1$, $p=2$, and $q=1$.
The result is shown in the top plot of Fig. \ref{fig:stacked_cost_function_j1}, where the black line indicates the solution manifold $\xi_d(\theta)$ and the red dot is the optimal solution $\xi_d(\theta^*)$.
The dashed line shows a reference direction $a_{ref}$ for comparison with the cost proposed in next section.

\subsection{Minimum norm with azimuth penalty}

In this case study, we aim to achieve minimum norm thrust allocation while additionally penalizing the azimuth variation.
It may seem challenging to minimize both the magnitude and the azimuth rate when working with the rectangular form of the thrust allocation.
Motivated by the work of \citet{mayhew2010hybrid} on potential functions on $\mathbb{S}^1$, we propose a convex cost function that can account for the azimuth angle and force magnitude to solve this problem.
Let $a_{ref,i} \in \mathbb{S}^1$ be a reference azimuth direction vector and $z \in \mathbb{R}^2$ be a rectangular force vector for a thruster.
The following cost function
\begin{align}
   P_i(z) \coloneqq w_i (|z| - \lambda_i a_{ref,i}^\top z), \quad w_i > 0, \ \lambda_i \in [0, 1],
   \label{04:potential_function_p}
\end{align}
has the property to minimize both the magnitude $|z|$ and relative difference in direction between $z$ and $a_{ref,i}$, using
\begin{align}
   \nabla_{z} P_i (z) = w_i \Bigg(\frac{z^\top}{|z|} - \lambda_i a_{ref,i}^\top\Bigg).
   \label{eq:03_p_i}
\end{align}
In the case of $\lambda_i=1$, the minimizer becomes $\frac{z^*}{|z^*|} = a_{ref,i}$, that is, the azimuth direction of $z$ aligns with $a_{ref,i}$.
In the case of $\lambda_i=0$ the minimum norm cost is captured.
Note that the gradient does not exist for $\nabla_z P_i(0)$; hence, a small regularization constant is included in the dynamic task assignment.
The applied cost function becomes
\begin{align}
   J_i(t, \theta) = P_i(\xi_{d,i}(t, \theta)), \ i \in \{1,\dots, m_1\},
\end{align}
for each azimuth thruster.

For the fixed-azimuth thrusters, the force $z$ is a scalar.
Consequently the direction $a_{ref,i}$ is also a scalar that reflects the choice between positive or negative force direction.
Letting $\sigma_{ref,i} \coloneqq \text{sgn}(F_{ref,i})$ where $F_{ref,i}$ is a reference force, we define a similar potential function to penalize change in the force direction for fixed-azimuth thrusters,
\begin{gather}
   L_i(z) \coloneqq w_i (|z| - \lambda_i \sigma_{ref,i} z), \quad w_i > 0, \ \lambda_i \in [0, 1], \\
   \nabla_{z} L_i = w_i \Big(\frac{z}{|z|} - \lambda_i \sigma_{ref,i}\Big). \label{eq:03_l_i}
\end{gather}
Constraining this along the solution manifold $\xi_d(t, \theta)$, we use the cost function
\begin{gather}
   J_i(t, \theta) = L_{i}(\xi_{d,i}(t, \theta)), \ i \in \{m_1 + 1,\dots, m\}
\end{gather}
for the fixed-azimuth thrusters.

For the dynamic task, we sum up the azimuth thrusters and fixed thrusters separately in the cost function,
\begin{align}
   J(t,\theta) \coloneqq \sum_{i=1}^{m_1} P_i(\xi_{d,i}(t,\theta)) +  \sum_{i=m_1+1}^{m} L_i(\xi_{d,i}(t,\theta)).
\end{align}
Recalling \eqref{eq:03_dynamic_assignment}, \eqref{eq:03_p_i}, and \eqref{eq:03_l_i}, the dynamic task is implemented by
\begin{align}
   \begin{aligned}
      \upsilon(t, \theta) :=  - & \gamma \Bigg[  \sum_{i=1}^{m_1} w_i Q_i ^\top \Big( \frac{\xi_{d,i}}{|\xi_{d,i}| + \varepsilon} - \lambda_i a_{ref,i}^\top \Big) \\
                                & +\sum_{i=m_1+1}^{m} w_i Q_i^\top \Big( \frac{\xi_{d,i}}{|\xi_{d,i}| + \varepsilon} - \lambda_i \sigma_{ref,i} \Big) \Bigg],      \\
      \label{eq:03_dynamic_assignment_01}
   \end{aligned}
\end{align}
where $\gamma > 0$, $\varepsilon > 0$ is a small regularization value, and arguments $(t, \theta)$ for $\xi_{d,i}$ and $(t)$ for $a_{ref,i}$ were left out for brevity.
Choosing $c_i=w_i$ for the CLF in \eqref{eq:03_clf_00}, and assign this to \eqref{eq:03_nominal_maneuvering_control_design}, we complete by assigning \eqref{eq:03_dynamic_assignment_01} to \eqref{eq:02_manuevering_control_law}.

Using the cost function $P(z)$, the optimal solution is pulled closer to the $a_{ref,i}$ for $z \mapsto P_i(z)$, as illustrated in the bottom plot of Fig. \ref{fig:stacked_cost_function_j1}.
The values were set to $w_i=1$, $\lambda_i=0.9$, $B=[1,2]$, $\tau_{cmd}=-6$ so that $n=1$, $p=2$, and $q=1$, as well as $a_{ref,i} = [\sqrt{2}, -\sqrt{2}]^\top$.


\section{Results}\label{sec:results}

\begin{figure}
   \centering
   \includegraphics[width=0.7\linewidth]{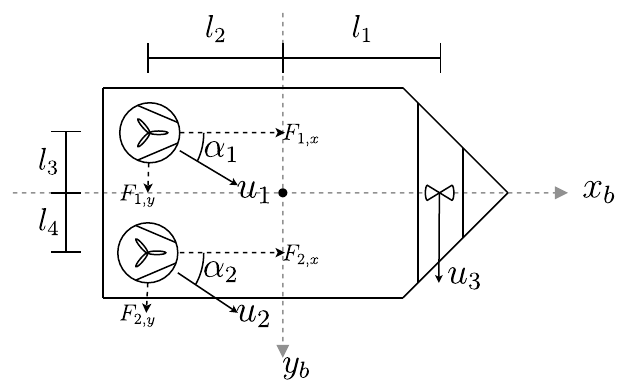}
   \caption{Thruster configuration of C/S Enterprise 1, with $l_1 = 0.3875$ m, $l_2 = 0.4574$ m, and $l_3 = l_4 = 0.055$ m.}
   \label{fig:05_actuator_configuration}
\end{figure}

We implemented the proposed algorithm and configured with the thruster configuration of the model vessel C/S Enterprise 1 \citep{skjetne2013recursive}.
The vessel has two Voith Schneider propellers at the stern and one tunnel thruster at the bow (see Fig. \ref{fig:05_actuator_configuration}).
For simplicity, we kept the actuator saturation limits at 1N.
We calculated $\dot{\tau}_{cmd}$ from $\tau_{cmd}$ numerically.
Note that the thruster dynamics are defined in \eqref{eq:03_control_input_phi}, and the presented algorithm only computes the allocated $F_i$ and $\alpha_i$ to be commanded to the local thruster control system.

\begin{figure}
   \centering
   \includegraphics[width=0.83\linewidth]{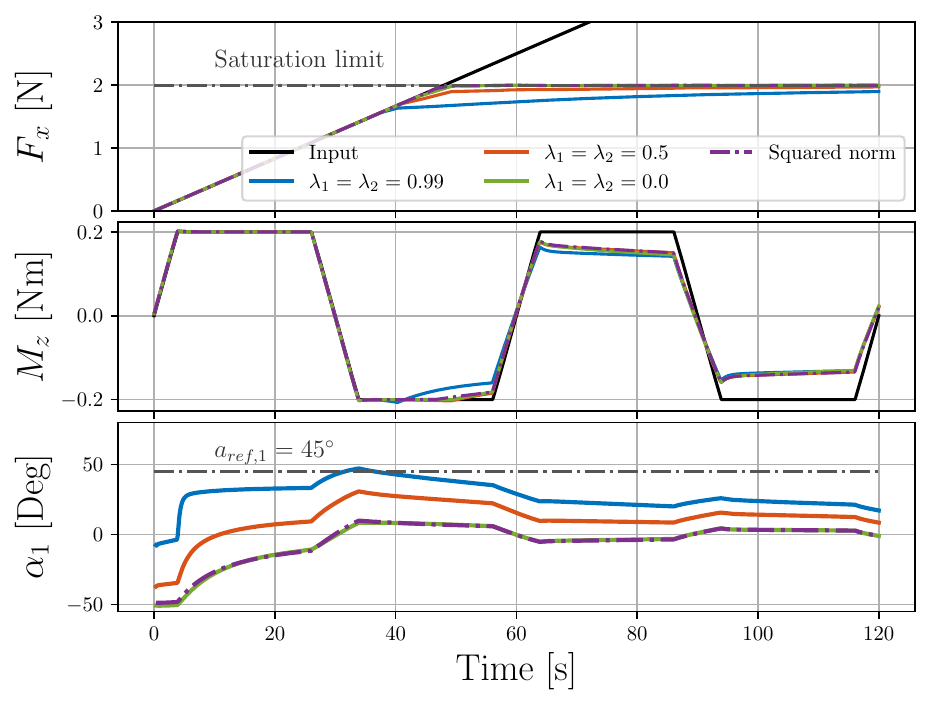}
   \caption{The plots compare outputs from two thrust allocators: \textit{minimum norm with azimuth penalty} (solid lines) and \textit{minimum squared norm} (dashed lines).Three $\lambda_i$ values are used: $\lambda_1=\lambda_2=0.99$, $\lambda_1=\lambda_2=0.5$, and $\lambda_1=\lambda_2=0.0$. Static azimuth reference angles are $\alpha_{ref,1}=45^\circ$ and $\alpha_{ref,2}=-45^\circ$. Only $\alpha_{1}$ is shown.}
   \label{fig:05_the_thing}
\end{figure}

First, we analyze the dynamic task performance of the \textit{minimum norm with azimuth penalty} reference filter.
In this case study, we ramped up $F_{cmd,x}$ steadily, while $F_{cmd,y}=0$, and oscillated $M_{cmd,z}$ between $0.2$ and $-0.2$ for the control commands and kept the configuration parameters the same as $\mu=0.1$, $\gamma=0.1$, $\rho_{i}=0.1$, and $\bar{\Omega}_i=1.0$.
Results are shown in Fig. \ref{fig:05_the_thing} where the system response for $F_x$, $M_z$, and azimuth angle for the port thruster, $\alpha_1$, are presented.
Fig. \ref{fig:05_the_thing} indicates how the proposed method forces the azimuth angle towards $\alpha_{ref,i}$, and this behavior becomes more apparent with increasing values for $\lambda_i$.
Additionally, the pure minimum norm case is recovered when $\lambda_i=0$, closely matching the \textit{minimum squared norm} thrust allocation output.

\begin{figure}
   \centering
   \includegraphics[width=0.83\linewidth]{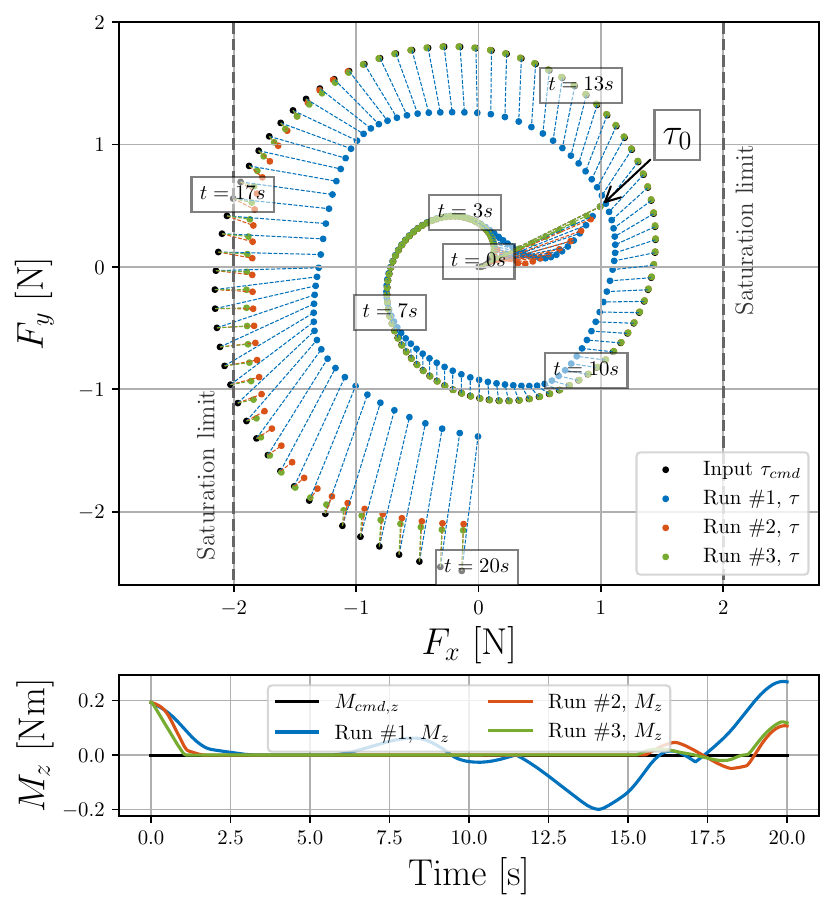}
   \caption{Thrust allocation commands and responses for test runs according to Table \ref{table:01_reference_filter_values}. The top plot compares the allocated thrust, shown as blue, red, and green dots, with $\tau_{cmd}$ shown as black dots, connected with lines to link corresponding $\tau_{cmd}$ values on the XY plane. The bottom plot shows the corresponding moment responses.}
   \label{fig:05_shell}
\end{figure}

\begin{figure}
   \centering
   \includegraphics[width=0.83\linewidth]{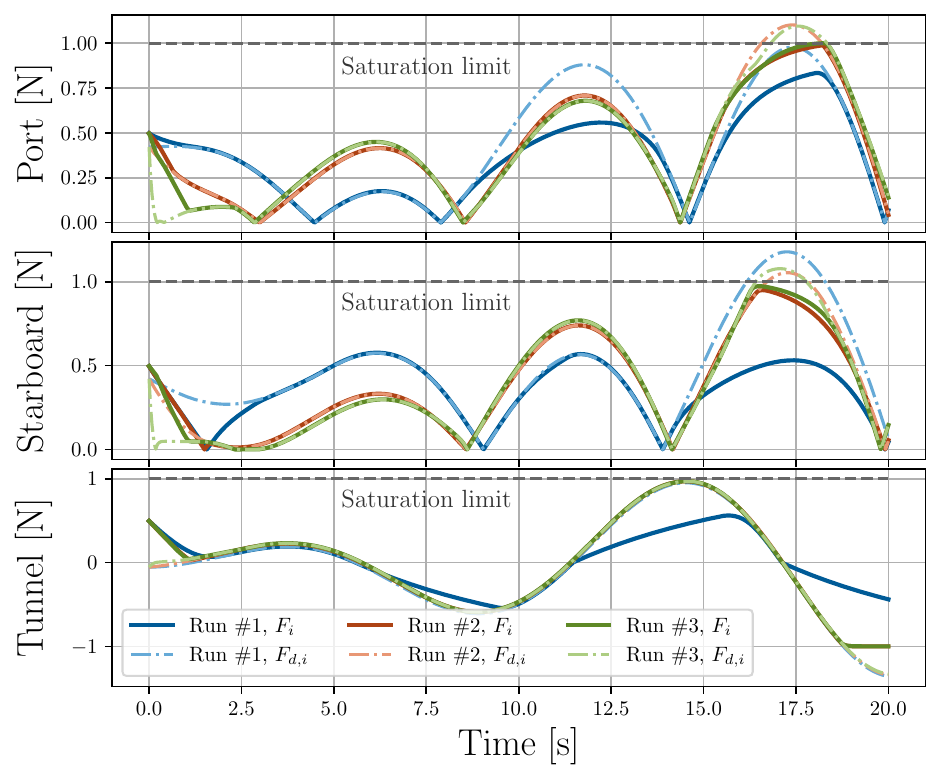}
   \caption{Thruster force outputs for different test runs according to Table \ref{table:01_reference_filter_values}, with $F_i = |\xi_i|$ and $F_{d,i} = |\xi_{d,i}|$ for $i=1,2$ and $F_3=\xi_3$, $F_{d,3} = \xi_{d,3}$ for the tunnel thruster, denoting the thruster forces and their desired values, respectively, for each run.}
   \label{fig:06_thruster_forces}
\end{figure}

\begin{figure}
   \centering
   \includegraphics[width=0.83\linewidth]{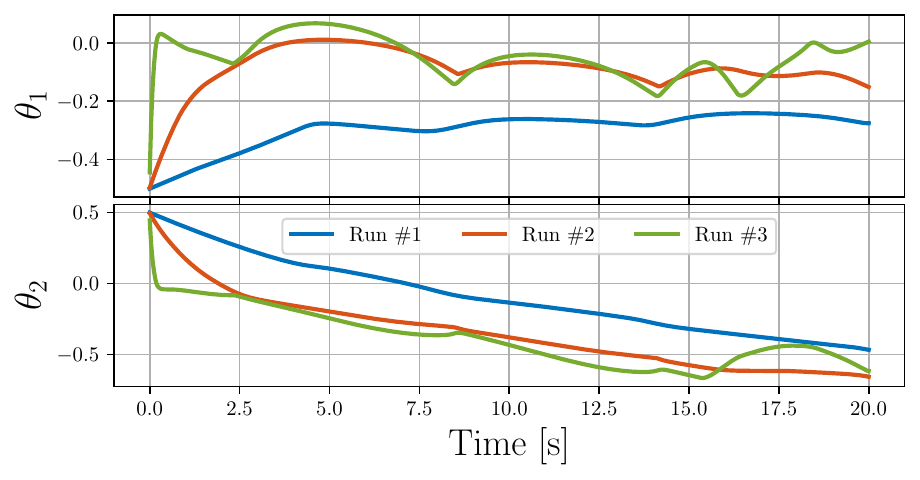}
   \caption{Reference filter parameter evolution over time for different test runs. The convergence of $\theta$ in the initial response is faster for higher $\mu$ values.}
   \label{fig:07_theta}
\end{figure}

We next analyze the effect of different configuration parameters on the \textit{minimum norm with azimuth penalty} thrust allocation with the CBF in \eqref{eq:03_cbf_abs_1}.
Figures \ref{fig:05_shell}, \ref{fig:06_thruster_forces}, and \ref{fig:07_theta} show the outputs of the test runs with values given in Table \ref{table:01_reference_filter_values} for the spiraling $\tau_{cmd}$ shown in Fig. \ref{fig:05_shell} represented with black dots.
We initialize the filter at $\xi_1(0)=[0.5, 0.0]^\top$, $\xi_2(0)=[0.5, 0.0]^\top$, $\xi_3(0)=0.5$, and $\theta(0) = [0.5, 0.5]^\top$ to better visualize the transient behavior of the filter.

\begin{table}[ht]
   \centering
   \caption{Values used in the test runs}
   \begin{tabular}{l|l|l|l|l|l|l|l|l}
      Run & $\gamma$ & $\mu$ & $\rho_i$ & $\zeta_i$ & $\lambda_i$ & $\bar{\Omega}_i$ & $\alpha_{ref,1}$ & $\alpha_{ref,2}$ \\ \hline \hline
      \#1 & 0.05     & 0.01  & 5.0      & 0.1       & 0.99        & 0.5              & $45^\circ$       & $-45^\circ$      \\ \hline
      \#2 & 0.1      & 0.5   & 0.1      & 0.01      & 0.99        & 0.5              & $45^\circ$       & $-45^\circ$      \\ \hline
      \#3 & 0.2      & 10.0  & 0.1      & 0.01      & 0.99        & 0.5              & $45^\circ$       & $-45^\circ$
   \end{tabular}
   \label{table:01_reference_filter_values}
\end{table}

In Fig. \ref{fig:05_shell}, the top plot shows the thrust allocation results on the XY-plane, where black dots represent $\tau_{cmd}$ and colored dots represent the allocated thrust with lines connecting the allocated outputs to the corresponding $\tau_{cmd}$.
The bottom plot shows the thrust allocation response to the commanded moment $M_{cmd,z}=0$.

The effect of the CBF time constants $\rho_i$ on the allocated thrust is clearly visible in figures \ref{fig:05_shell} and \ref{fig:06_thruster_forces}, particularly when comparing the response of Run \#1, where the $\rho_i$ value is the highest, to runs \#2 and \#3.
As $\rho_i$ increases, the system becomes more conservative when $\xi_i$ approaches the saturation limits of the thrusters.
Additionally, the system responses for runs \#2 and \#3 demonstrate that higher $\mu$ and $\gamma$ values lead to faster convergence of $\theta$, as illustrated in Fig. \ref{fig:07_theta}, where $\mu$ gives the transient behavior at the start of the response to rapidly minimize $\theta \mapsto V(t, \theta, \xi)$ and thereafter $\gamma$ ensures the minimization of $\theta \mapsto J(t, \theta)$.

\section{Conclusion}\label{sec:conclusions}

In this study, we proposed an alternative approach to solving the thrust allocation problem using maneuvering theory to design a dynamic thrust allocation reference filter.
The proposed method solves the geometric task of making the virtual force vector of the reference filter converge to and stay on a solution manifold for the effector model.
For this purpose, we employed a control Lyapunov function to design a nominal control that respects the force rate limits of the thrusters, and a control barrier function was employed to ensure the force saturations of the thrusters.
We further showed how a dynamic task can be designed to achieve optimal thrust allocation by minimizing a cost function as a dynamic assignment, and we presented two case studies.
The first showed how to solve the common minimum squared norm thrust allocation, whereas the second also included an azimuth penalty.
Finally, we presented our numerical results to show the capabilities of the algorithm.

\begin{ack}
   We thank Stian André Elvenes for his contributions to testing our ideas in his master thesis.
\end{ack}

\bibliography{ifacconf}             

\begin{thebibliography}{12}
\providecommand{\natexlab}[1]{#1}
\providecommand{\url}[1]{\texttt{#1}}
\providecommand{\urlprefix}{URL }
\expandafter\ifx\csname urlstyle\endcsname\relax
  \providecommand{\doi}[1]{doi:\discretionary{}{}{}#1}\else
  \providecommand{\doi}{doi:\discretionary{}{}{}\begingroup
  \urlstyle{rm}\Url}\fi

\bibitem[{Aguiar et~al.(2008)Aguiar, Hespanha, and
  Kokotovi{\'c}}]{aguiar2008performance}
Aguiar, A.P., Hespanha, J.P., and Kokotovi{\'c}, P.V. (2008).
\newblock Performance limitations in reference tracking and path following for
  nonlinear systems.
\newblock \emph{Automatica}, 44(3).

\bibitem[{Encarna{\c{c}}ao and Pascoal(2001)}]{encarnaccao2001combined}
Encarna{\c{c}}ao, P. and Pascoal, A. (2001).
\newblock Combined trajectory tracking and path following: an application to
  the coordinated control of autonomous marine craft.
\newblock In \emph{Proc. 40th IEEE Conf. Decision Control}.

\bibitem[{Fossen(2011)}]{fossen2011handbook}
Fossen, T.I. (2011).
\newblock \emph{Handbook of marine craft hydrodynamics and motion control}.
\newblock John Wiley \& Sons.

\bibitem[{Hauser and Hindman(1995)}]{hauser1995maneuver}
Hauser, J. and Hindman, R. (1995).
\newblock Maneuver regulation from trajectory tracking: Feedback linearizable
  systems.
\newblock \emph{IFAC Proc. Volumes}, 28(14).

\bibitem[{Johansen and Fossen(2013)}]{johansen2013control}
Johansen, T.A. and Fossen, T.I. (2013).
\newblock Control allocation—a survey.
\newblock \emph{Automatica}, 49(5).

\bibitem[{Marley et~al.(2021)Marley, Skjetne, Basso, and
  Teel}]{marley2021maneuvering}
Marley, M., Skjetne, R., Basso, E., and Teel, A.R. (2021).
\newblock Maneuvering with safety guarantees using control barrier functions.
\newblock \emph{IFAC-PapersOnLine}, 54(16).

\bibitem[{Mayhew and Teel(2010)}]{mayhew2010hybrid}
Mayhew, C.G. and Teel, A.R. (2010).
\newblock Hybrid control of planar rotations.
\newblock In \emph{Proc. 2010 Am. Control Conf.} IEEE.

\bibitem[{Peng et~al.(2016)Peng, Wang, and Wang}]{peng2016containment}
Peng, Z., Wang, J., and Wang, D. (2016).
\newblock Containment maneuvering of marine surface vehicles with multiple
  parameterized paths via spatial-temporal decoupling.
\newblock \emph{IEEE/ASME Trans. on Mechatronics}, 22(2).

\bibitem[{Skjetne(2005)}]{skjetne2005maneuvering}
Skjetne, R. (2005).
\newblock The maneuvering problem.
\newblock \emph{NTNU, PhD-thesis, Norwegian University of Science \&
  Technology, Trondheim, Norway.}

\bibitem[{Skjetne et~al.(2011)Skjetne, J{\o}rgensen, and
  Teel}]{skjetne2011line}
Skjetne, R., J{\o}rgensen, U., and Teel, A.R. (2011).
\newblock Line-of-sight path-following along regularly parametrized curves
  solved as a generic maneuvering problem.
\newblock In \emph{50th Conf. on Decision and Control and European Control
  Conf.}

\bibitem[{Skjetne and Kjerstad(2013)}]{skjetne2013recursive}
Skjetne, R. and Kjerstad, {\O}.K. (2013).
\newblock Recursive nullspace-based control allocation with strict
  prioritization for marine craft.
\newblock \emph{IFAC Proc. Volumes}, 46(33).

\bibitem[{Xu et~al.(2015)Xu, Tabuada, Grizzle, and Ames}]{xu2015robustness}
Xu, X., Tabuada, P., Grizzle, J.W., and Ames, A.D. (2015).
\newblock Robustness of control barrier functions for safety critical control.
\newblock \emph{IFAC-PapersOnLine}, 48(27).

\end{thebibliography}

\appendix
\end{document}